\documentclass[]{tMOP2e}
\pdfoutput=1
\usepackage{psfrag,graphicx}
\usepackage{float}
\usepackage{graphicx}
\usepackage{nccmath}
\usepackage{psfrag}
\usepackage{amsmath}
\usepackage{times}
\usepackage{hhline}
\def\be{\begin{equation}}
\def\ee{\end{equation}}
\def\ba{\begin{array}}
\def\ea{\end{array}}
\def\bea{\begin{eqnarray}}
\def\eea{\end{eqnarray}}
\def\no{\nonumber}

\def\sech{\rm{sech}}

\begin{document}

\articletype{Article}

\title{Semirational and symbiotic self-similar rogue waves in a (2+1)-dimensional graded-index waveguide}
\author{Kanchan Kumar De$^{a}$, Thokala Soloman
Raju$^{b}$$^{\ast}$\thanks{$^\ast$Corresponding author. Email:
solomonr\textunderscore{thokala@yahoo.com}}, C. N. Kumar$^{c}$,
and Prasanta K. Panigrahi$^{d}$
\\\vspace{6pt}
$^{a}${\em{Department of Physics, M.Sc. College,  Saharanpur,
Uttar Pradesh 247001 India}};
 $^{b}${\em{Indian Institute of Science Education and Research
(IISER) Tirupati, Andhra Pradesh 517507, India}};
 $^{c}${\em{Department of Physics, Panjab University, Chandigarh 160014, India}};
 $^{d}${\em{Indian Institute of Science Education and Research
(IISER) Kolkata,
 Mohanpur, Nadia 741246, India}}\\
 }

\maketitle

\begin{abstract} We have investigated the ($2+1$)-dimensional variable
coefficient coupled nonlinear Schr\"{o}dinger equation (vc-CNLSE)
in a graded-index waveguide. Similarity transformations are used
to convert the vc-CNLSE into constant coefficient CNLSE. Under
certain functional constraints we could extract semirational,
multi-parametric solution of the associated Manakov system. This
family of solutions include known Peregrine soliton, mixture of
either bright soliton and rogue wave or dark soliton and rogue
wave or breather and rogue wave. Under a distinct set of
self-phase modulation (SPM) and cross-phase modulation (XPM)
coefficients we could establish symbiotic existence of different
soliton pairs as solutions. These soliton pairs may constitute of
one bright and a dark soliton, two bright solitons or two dark
solitons. Finally, when two wave components are directly
proportional, we find bright and dark similaritons, self-similar
breathers and rogue waves as different solutions.
\end{abstract}
\section{Introduction}The mathematical model governing
wave propagation in graded-index nonlinear waveguides is the
variable coefficient coupled nonlinear Schr\"{o}dinger equation
(vc-CNLSE) \cite{Agrawal}. Apart from this, CNLSE is one of the
basic mathematical model for many physical systems. It governs the
wave propagation of two overlapped wave packets in hydrodynamics
\cite{Dhar}, Langmuir and acoustic waves in plasma \cite{Som},
Rossby waves \cite{Sun} etc. CNLSE also describes wave propagation
in twin core optical fibers \cite{Huanag}, birefringent fibers
\cite{Menyuk} and directional couplers \cite{Boumaza}.\\ We shall
consider ($2+1$)-dimensional vc-CNLSE in which dispersion,
nonlinearity, gain and tapering function have spatial dependence.
Kruglov et al. has given self-similar asymptotic solutions for
equivalent ($1+1$)-dimensional equation with constant coefficients
\cite{Kruglov}. This ($1+1$)-dimensional equation in the absence
of the intermodal dispersion term and with the equal strength of
SPM and XPM had been investigated by Tian et al. \cite{Tian} and
found one- and two-soliton solutions using the Darboux method.
Cardoso et al. \cite{Cardoso} found stable solutions for a similar
nonautonomous ($1+1$)-dimensional coupled NLSE.\\
Few stable localized solutions of the two dimensional NLSE with
distributed coefficients have been obtained by numerical
simulations \cite{Alexandrescu,Xu,Dai1}. Recently, propagation
behavior of two-breather, Kuznetsov-Ma soliton solutions were
studied based on ($2+1$)-dimensional coupled NLSE \cite{Dai2}.\\
In this article, we have found different semirational and
symbiotic solutions for a ($2+1$)-dimensional variable coefficient
coupled nonlinear Schr\"{o}dinger equation (vc-CNLSE). This
coupled equation is modelled for a graded-index waveguide.
Similarity transformations are used to convert the coupled
equations into constant co-efficient CNLSE. Under certain
functional constraints we could extract semirational,
multi-parametric solution of the associated Manakov system. This
family of solutions include known Peregrine soliton, mixture of
bright soliton and rogue wave, mixture of  dark soliton and rogue
wave, mixture of breather soliton and rogue wave. Also, under a
distinct set of self-phase modulation (SPM) and cross-phase
modulation (XPM) coefficients we could establish symbiotic
existence of different soliton pairs as solutions. These soliton
pairs may constitute of one bright and a dark soliton, two bright
solitons or two dark solitons. Finally, for a special
circumstance, when two wave components are directly proportional;
we find bright and dark similaritons, self-similar breathers and
rogue waves as possible solutions. Recently, some of the present
authors have also studied symbiotic multimode spatial similaritons
and rogons in inhomogeneously coupled optical fibers
\cite{soloman}. While we discuss the dynamics of these symbiotic
solitons in the next sections, here we mention the seminal works
done by Serkin and his group \cite{serkin11,serkin12,serkin13}.
\section{Model equation and similarity transformation} Wave propagation in inhomogeneous nonlinear waveguide can
be described by the vc-CNLSE \be\label{1}i\frac{\partial
u}{\partial z}+\frac{\beta(z)}{2}(\frac{\partial^{2}u}{\partial
x^{2}}+\frac{\partial^{2}u}{\partial
y^{2}})+\chi(z)(r_{11}|u|^{2}+r_{12}|v|^{2} )u
+\frac{1}{2}f(z)(x^{2}+y^{2})u = i~g(z)u\ee
\be\label{2}i\frac{\partial v}{\partial
z}+\frac{\beta(z)}{2}(\frac{\partial^{2}v}{\partial
x^{2}}+\frac{\partial^{2}v}{\partial
y^{2}})+\chi(z)(r_{21}|u|^{2}+r_{22}|v|^{2} )v
+\frac{1}{2}f(z)(x^{2}+y^{2})v = i~g(z)v\ee where $u(x,y,z)$ and
$v(x,y,z)$ are the the two normalized orthogonal components of
electric fields. $x,y$ are the dimensionless transverse variables
and z is the dimensionless propagation distance. Here, $\beta(z),
\chi(z), g(z)$ and $f(z)$ are the dispersion, nonlinearity, gain
and tapering function respectively. $r_{11}$ and $r_{22}$ are the
self-phase modulation (SPM) coefficients for $u(x,y,z)$ and
$v(x,y,z)$; $r_{21}$ and $r_{12}$ are the cross-coupling
coefficients which determine the strength of cross-phase
modulation (XPM). The refractive index distribution inside the
waveguide is assumed as $ n = n_0 + n_1 f(z)(x^2+y^2)+
n_2~\chi(z)I(x,y,z)$, $I(x,y,z)$ being the optical intensity.
Here, the first two terms represent the linear part of the
refractive index and the last term describes a Kerr-type
nonlinearity of the waveguide.

We employ gauge and similarity transformations \be\label{3}u =
A(z) e^{\Omega(z)} ~\Psi_{1}[X(x,y,z),\zeta(z)]
\exp[i\varphi(x,y,z)],\ee \be\label{4}v = A(z) e^{\Omega(z)}~
\Psi_{2}[X(x,y,z),\zeta(z)] \exp[i\varphi(x,y,z)]\ee in Equation
(\ref{1}) and (\ref{2}). Similarity variable and phase are assumed
as

\be\label{5}X(x,y,z) = \frac{k x + l y - X_{c}(z)}{W(z)},\ee
\be\label{6}\varphi(x,y,z) = a(z)(x^{2}+ y^{2})+ b(z)(\frac{x}{k}+
\frac{y}{l})+ c(z),\ee where $W(z)$ and $X_{c}(z)$ are the
dimensionless beam width and position coordinate of the
self-similar wave center. $a(z),b(z)$ and $c(z)$ are the
parameters related to phase-front curvature, frequency shift and
phase offset respectively. When the system satisfy the following
constraints \be\label{7}\Omega(z) = \int_{0}^{z}g(t) dt,\ee
\be\label{8}A(z) = \frac{A_{0}}{W},\ee \be\label{9}\chi(z) =
\frac{\beta(k^{2}+l^{2})}{A_{0}^{2}} e^{-2\Omega(z)},\ee
\be\label{10}f(z) = \frac{\beta
W_{zz}-\beta_{z}W_{z}}{\beta^{2}W},\ee \be\label{11}X_{c}(z) =2 m
W \int_{0}^{z}\frac{\beta(t)}{W^{2}(t)}dt,\ee\be\label{12}\zeta(z)
= \int_{0}^{z}\frac{(k^{2}+l^{2})\beta(t)}{W^{2}(t)}dt,\ee and
\be\label{13}a(z)=\frac{W_{z}}{2 \beta W},~b(z)= \frac{m}{W},~c(z)
=- m^{2} (\frac{1}{2 k^{2}}+\frac{1}{2 l^{2}})
\int_{0}^{z}\frac{\beta(t)}{W^{2}(t)}dt,\ee with $A_{0},k,l,m$
being real constants; Equation (\ref{1}) and (\ref{2}) reduces
respectively to
\be\label{14}i\frac{\partial\Psi_{1}}{\partial\zeta}+\frac{1}{2}\frac{\partial^{2}\Psi_{1}}{\partial
X^{2}}+(r_{11} |\Psi_{1}|^{2}+ r_{12} |\Psi_{2}|^{2})\Psi_{1}=0\ee
and
\be\label{15}i\frac{\partial\Psi_{2}}{\partial\zeta}+\frac{1}{2}\frac{\partial^{2}\Psi_{2}}{\partial
X^{2}}+(r_{21} |\Psi_{1}|^{2}+ r_{22} |\Psi_{2}|^{2})\Psi_{2}=0\ee
In the subsequent analysis we shall assume $g(z)= \tanh
z;~A_{0}=1;~k=l=2;~m=1;~\beta=1; y=0$. We also assume the form of
the tapering function $f(z)=\frac{1}{\beta}(1-2~\mbox{sech}^2z)$
and hence, from Equation (\ref{10}) we have $W(z)=\mbox{sech}z$.

\section{Manakov system and semirational solutions}
A Manakov system is the one where the values of self-phase
modulation (SPM) and cross-phase modulation (XPM) are same in a
coupled NLSE. Setting $r_{i j} = 1$, ($i, j = 1, 2$) in Equation
(\ref{14}) and (\ref{15}) we have the associated Manakov system
given by
\be\label{16}i\frac{\partial\Psi_{1}}{\partial\zeta}+\frac{1}{2}\frac{\partial^{2}\Psi_{1}}{\partial
X^{2}}+( |\Psi_{1}|^{2}+  |\Psi_{2}|^{2})\Psi_{1}=0\ee and
\be\label{17}i\frac{\partial\Psi_{2}}{\partial\zeta}+\frac{1}{2}\frac{\partial^{2}\Psi_{2}}{\partial
X^{2}}+( |\Psi_{1}|^{2}+  |\Psi_{2}|^{2})\Psi_{2}=0\ee

This Manakov system represented by coupled Equation (\ref{16}) and
(\ref{17}) possesses the multiparametric soliton/rogue wave
solutions \cite{Baronio}, given by \be\label{18} \left(%
\begin{array}{c}
  \Psi_{1} \\
  \Psi_{2} \\
\end{array}%
\right)= e^{i\sigma\zeta}\left[\frac{L}{N}\left(%
\begin{array}{c}
  a_{1} \\
  a_{2} \\
\end{array}%
\right)+\frac{M}{N}\left(%
\begin{array}{c}
  a_{2} \\
  -a_{1} \\
\end{array}%
\right)\right],\ee with $ L=\frac{3}{2}-2\sigma^{2}\zeta^{2}-2
\sigma X^{2}+4i\sigma\zeta +|h|^{2}e^{2aX}, M=4
h(aX-i\sigma\zeta-\frac{1}{2})e^{(aX+i\frac{\sigma}{2}\zeta)}$ and
$N= \frac{1}{2}+2 \sigma^{2}\zeta^{2}+2\sigma X^{2} +
|h|^{2}e^{2aX}$. Here, $a_{1}$ and $a_{2}$ are arbitrary real
parameters. Actually, $a_{1}$ and $a_{2}$ are the amplitudes of
the background plane waves. $h$ is a complex arbitrary constant
and $\sigma = a^{2}=a^{2}_{1} + a^{2}_{2}$. These solutions have
both exponential and rational dependence on coordinates and,
therefore, they are called semirational. For special parameter
values, these solutions produce known rogue waves such as the
Peregrine soliton, mixture of bright soliton and rogue wave or mixture of dark soliton and rogue wave.\\
The corresponding expressions of intensity can be written as
\be\label{19} \left(%
\begin{array}{c}
  I_{u} \\
  I_{v} \\
\end{array}%
\right)=\frac{A_{0}^{2} e^{2\Omega}}{W^2} \left[\frac{|L|^{2}}{|N|^{2}}\left(%
\begin{array}{c}
  a^{2}_{1} \\
  a^{2}_{2} \\
\end{array}%
\right)+\frac{2Re(LM^{*})}{|N|^{2}}\left(%
\begin{array}{c}
 a_{1} a_{2} \\
 - a_{1}a_{2} \\
\end{array}%
\right)+\frac{|M|^{2}}{|N|^{2}}\left(%
\begin{array}{c}
  a^{2}_{2} \\
  a^{2}_{1} \\
\end{array}%
\right)\right].\ee
\begin{figure}[h*]
\includegraphics[scale=0.7]{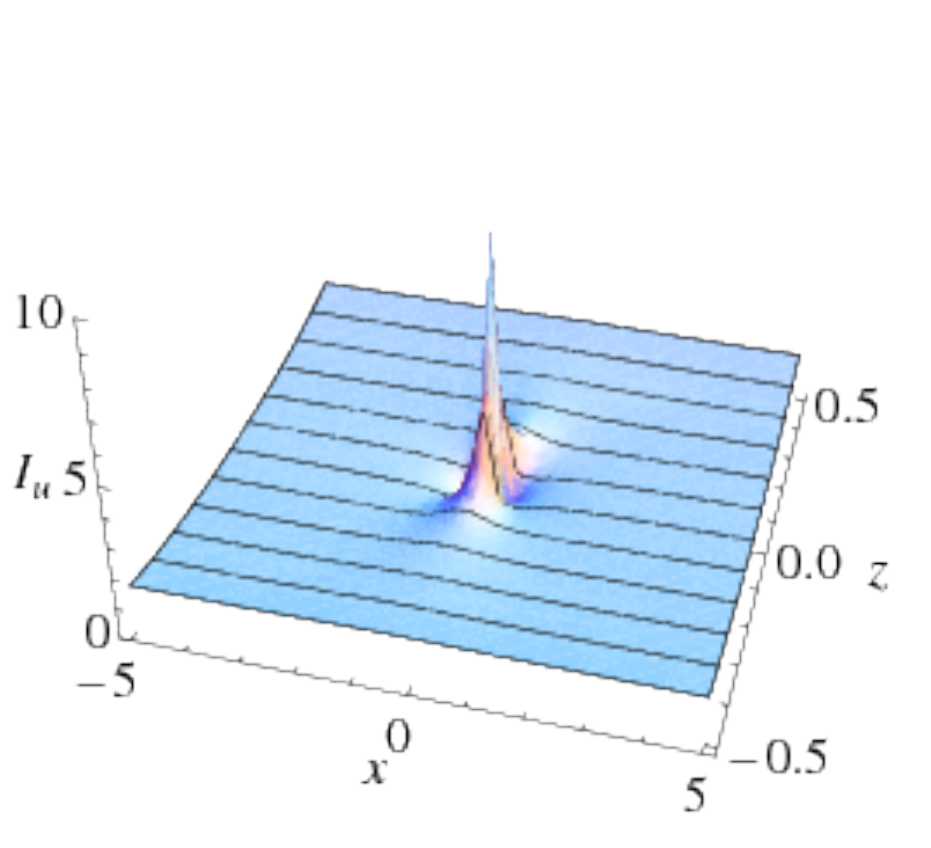}~~~~~\includegraphics[scale=0.7]{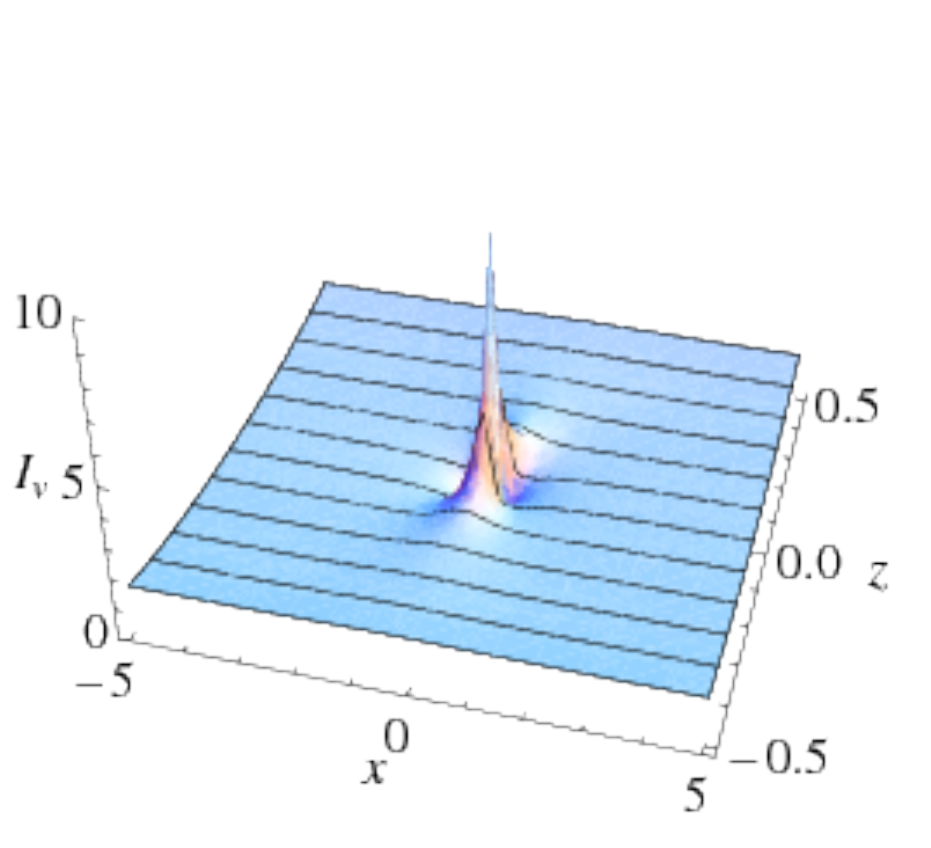}
\caption{\label{fig1}Freak waves intensity patterns $I_{u}$ and
$I_{v}$ of Equation (\ref{19}). Here, $a_{1}=1, a_{2}=1, h=0$.}
\end{figure}
\begin{figure}[h*]
\includegraphics[scale=0.7]{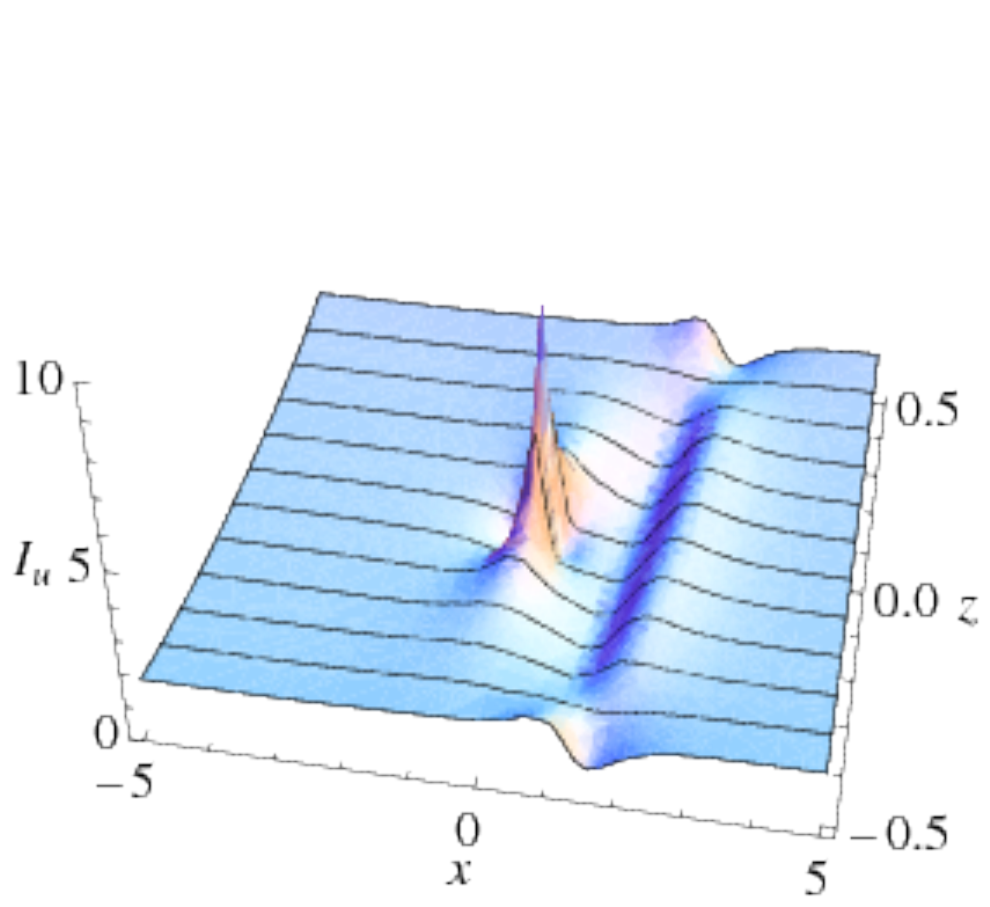}~~~~~\includegraphics[scale=0.7]{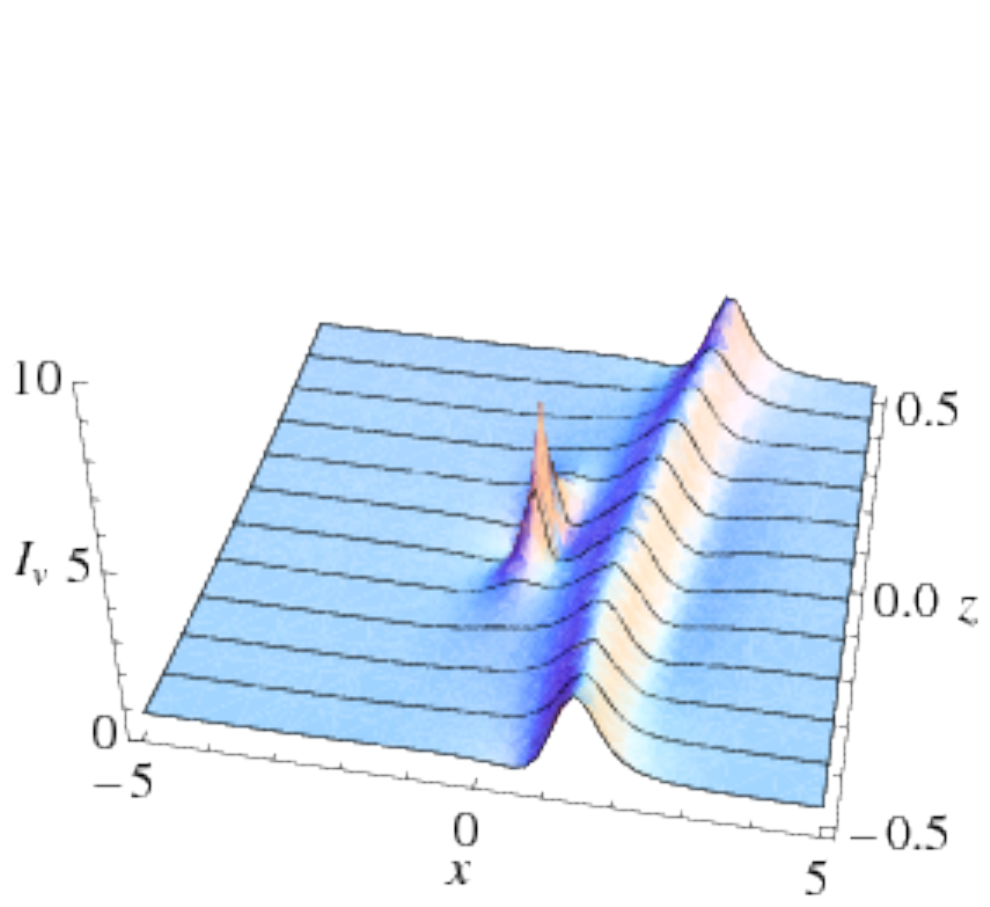}
\caption{\label{fig2}Intensity patterns $I_{u}$ and $I_{v}$ for
$a_{1}=1, a_{2}=0.6, h=0.1$. }
\end{figure}
\begin{figure}[h*]
\includegraphics[width=0.5\textwidth,natwidth=310,natheight=242]{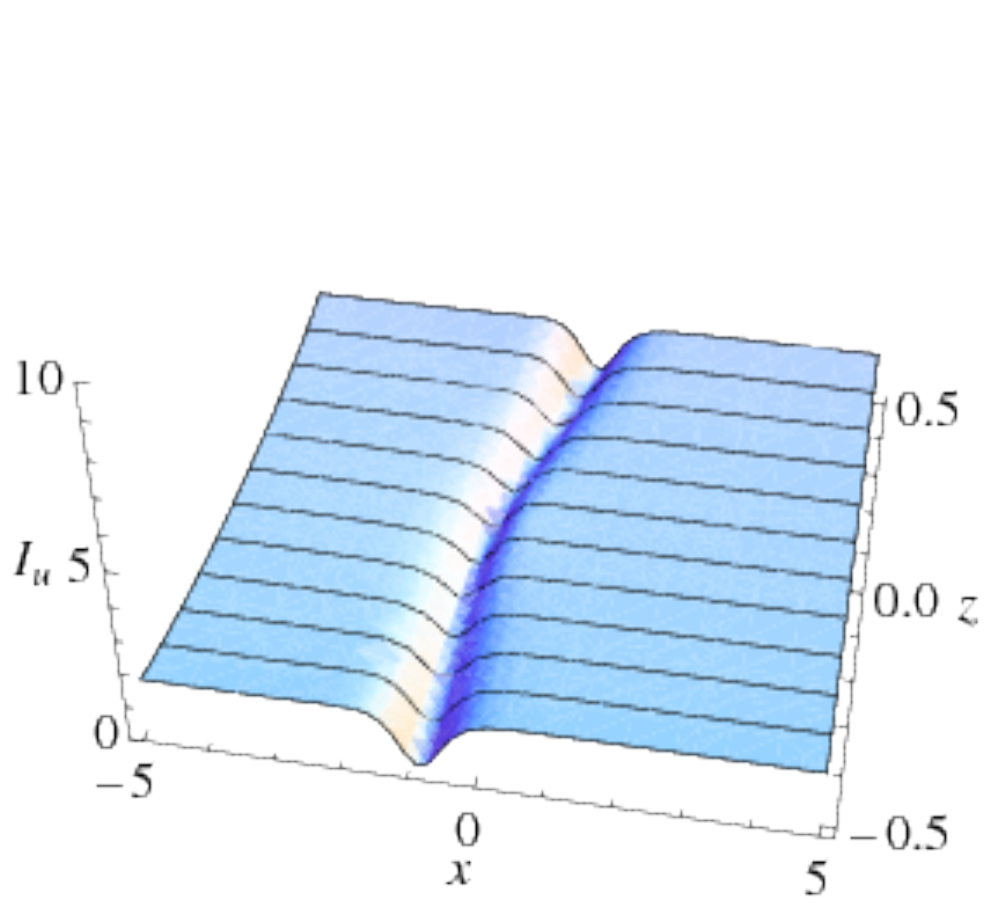}\includegraphics[width=0.5\textwidth,natwidth=310,natheight=242]{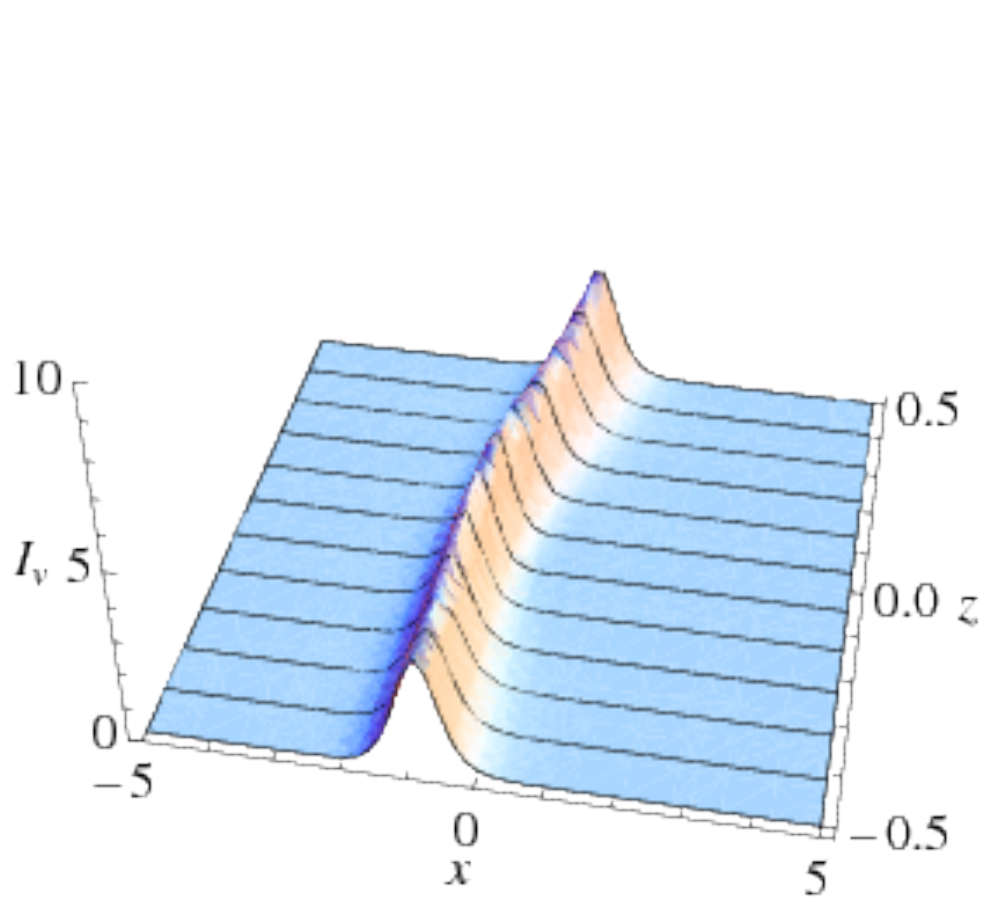}
\caption{\label{fig3}Intensity patterns $I_{u}$ and $I_{v}$ for
$a_{1}=1, a_{2}=0, h=15$. }
\end{figure}
\begin{figure}[h*]
\includegraphics[width=0.5\textwidth,natwidth=310,natheight=242]{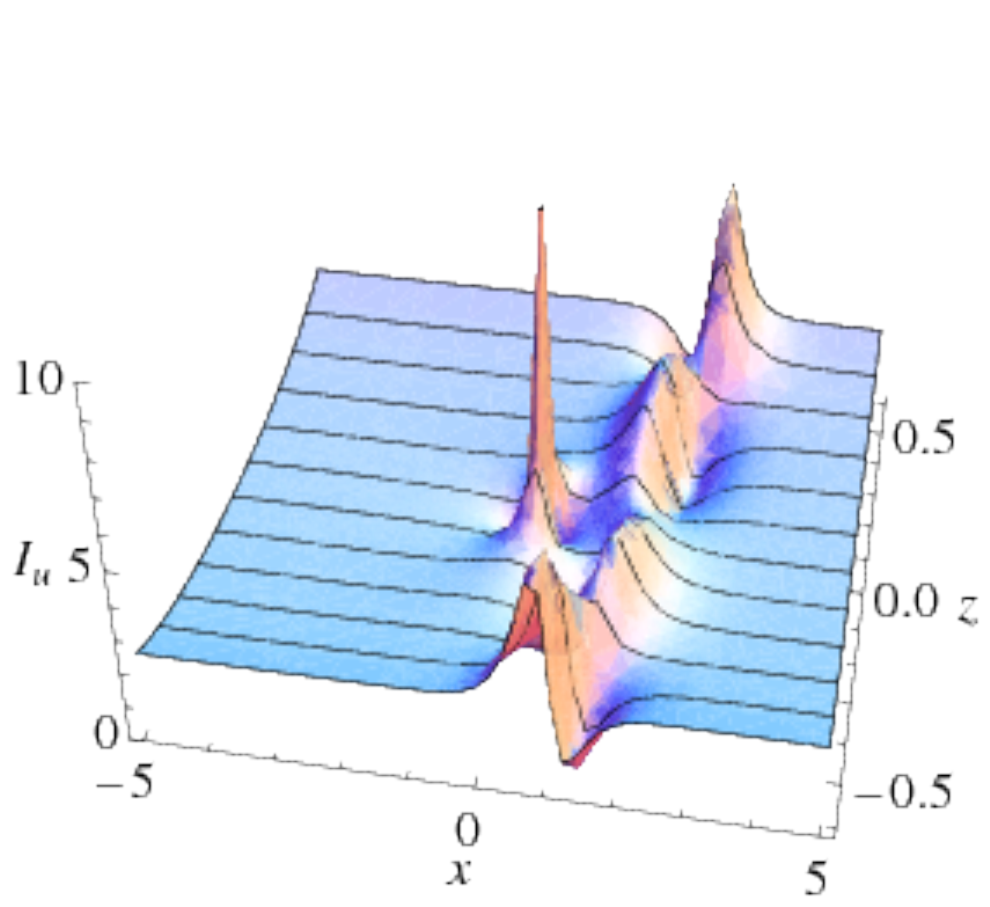}\includegraphics[width=0.5\textwidth,natwidth=310,natheight=242]{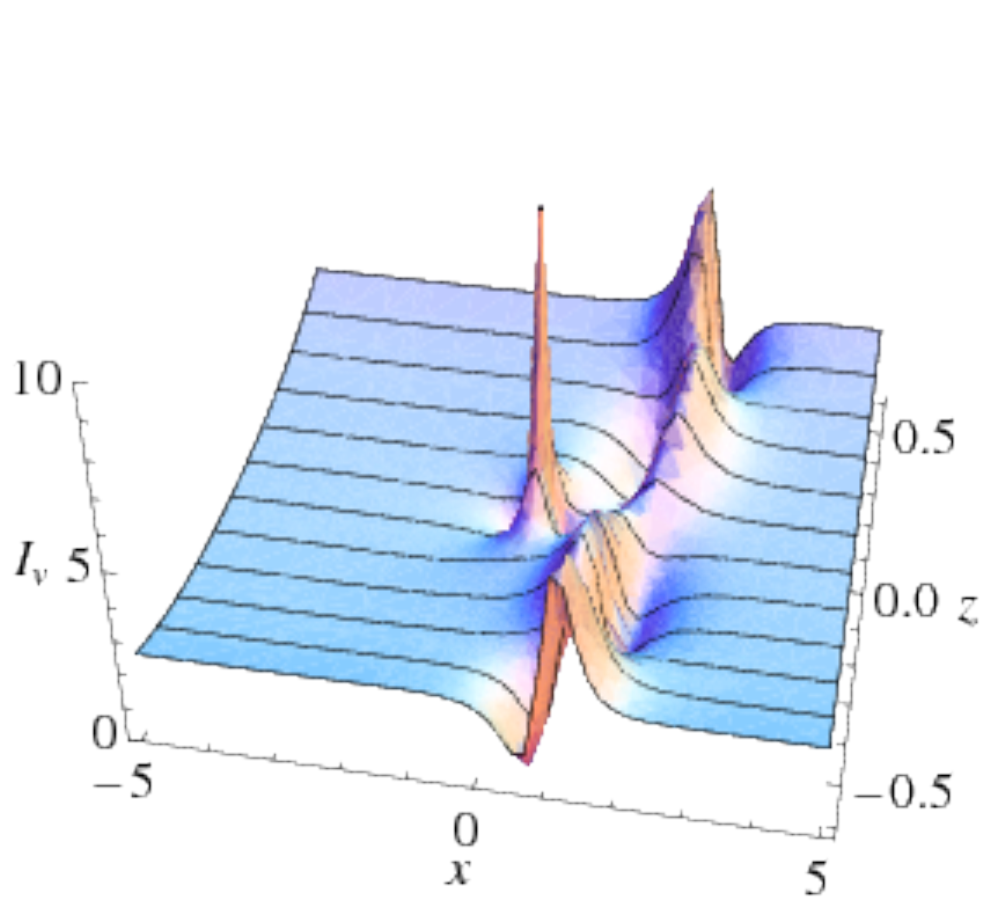}
\caption{\label{fig4}Intensity patterns $I_{u}$ and $I_{v}$ for
$a_{1}=1.1, a_{2}=1.1, h=\frac{i}{15}$. }
\end{figure}
\begin{figure}[h*]
\begin{center}
\includegraphics[width=0.5\textwidth,natwidth=310,natheight=242]{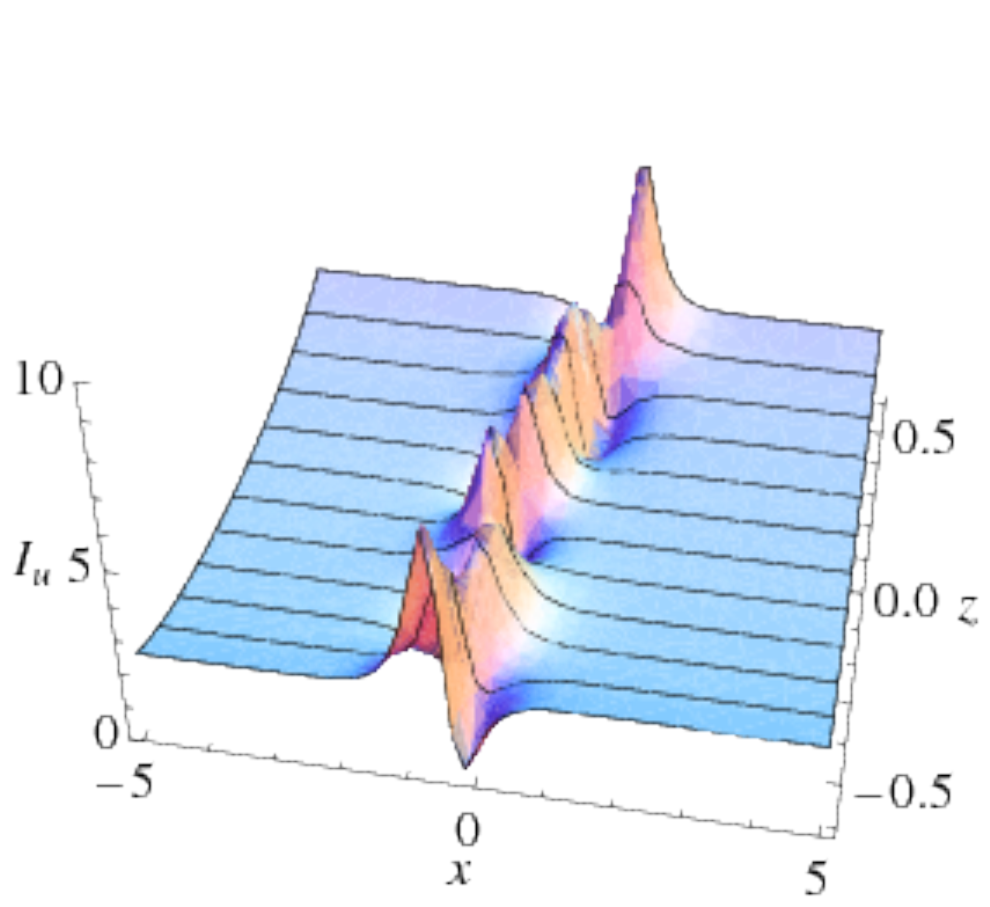}\includegraphics[width=0.5\textwidth,natwidth=310,natheight=242]{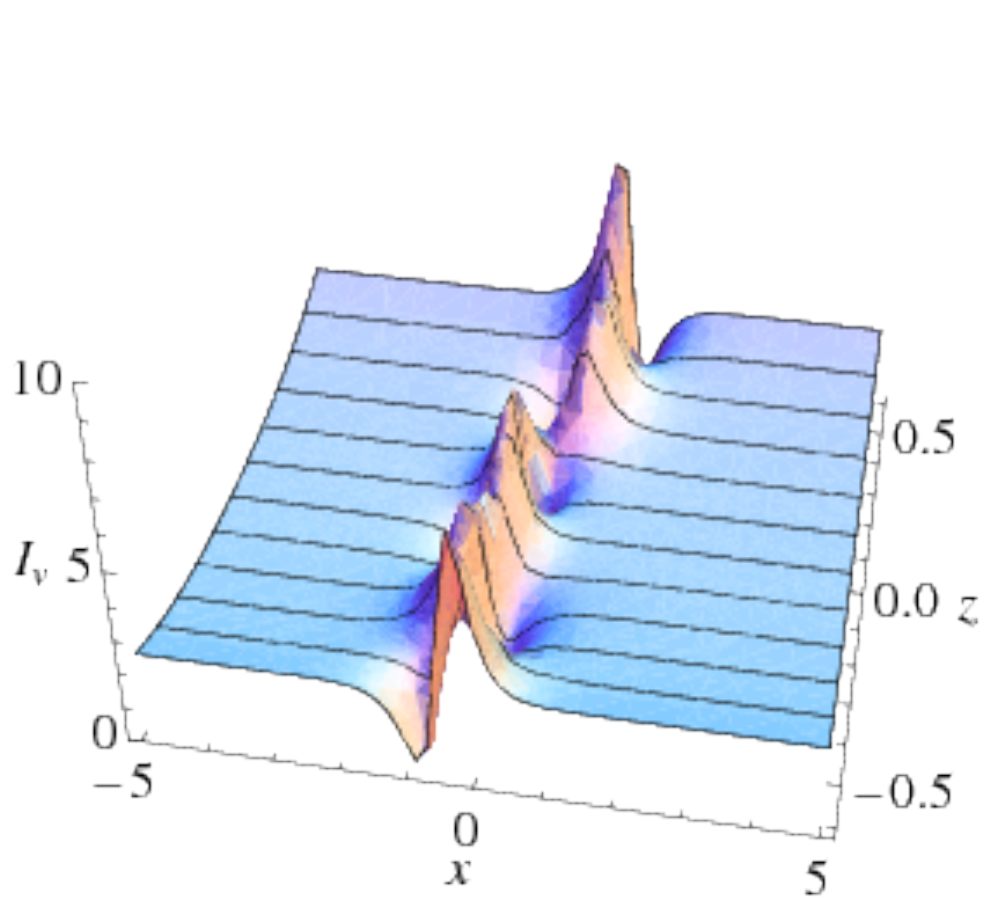}
\caption{\label{fig5}Intensity patterns $I_{u}$ and $I_{v}$ for
$a_{1}=1.1, a_{2}=1.1, h=15 i$. } \end{center}
\end{figure}
Setting $h=0$ implies $M=0$; in this particular case the
expression (18) yields the rational Peregrine solution
\cite{Peregrine,Bludov}. In this case $\Psi_{1}$ is merely
proportional to $\Psi_{2}$. In Figure \ref{fig1} we note that the
intensity is peaked at $x=0$ and $z=0$ with the maximum value $9
a_{j}$.\\ Each wave component $\Psi_{j}$ is generically a mixture
of a dark and a bright pulse. The superposition of the dark and
bright contributions in each of the two wave components may give
rise to complicated pulses. For small values of $|h|$, Peregrine
and dark/bright solitons coexists separately as shown in Figure
\ref{fig2}. By increasing $|h|$, Peregrine and dark/bright
solitons merge as shown in Figure \ref{fig3}. In this case,
Peregrine bump cannot be identified and the resulting bright/dark
pulse appear as a boomeron-type soliton, i.e., a soliton solution
with time dependent velocity \cite{Degasperis,Conforti}. In
general one can infer that, if $h\neq0$ the Peregrine bump
coexists with bright and dark solitons.\\ If all parameters $h,
a_{1}, a_{2}$ are nonvanishing, Equation (\ref{18}) describes the
dynamics of a breather-like wave along with Peregrine soliton.
Again, for small values of $|h|$ Peregrine and breather-like
soliton coexists separately as shown in Figure \ref{fig4}. For
increased $|h|$ value, Peregrine and breather soliton merge, as
shown in Figure \ref{fig5}.
\subsection{Generalized tapering and freak wave intensity}
\begin{figure}[h*]
\begin{center}
\includegraphics[width=0.5\textwidth,natwidth=310,natheight=242]{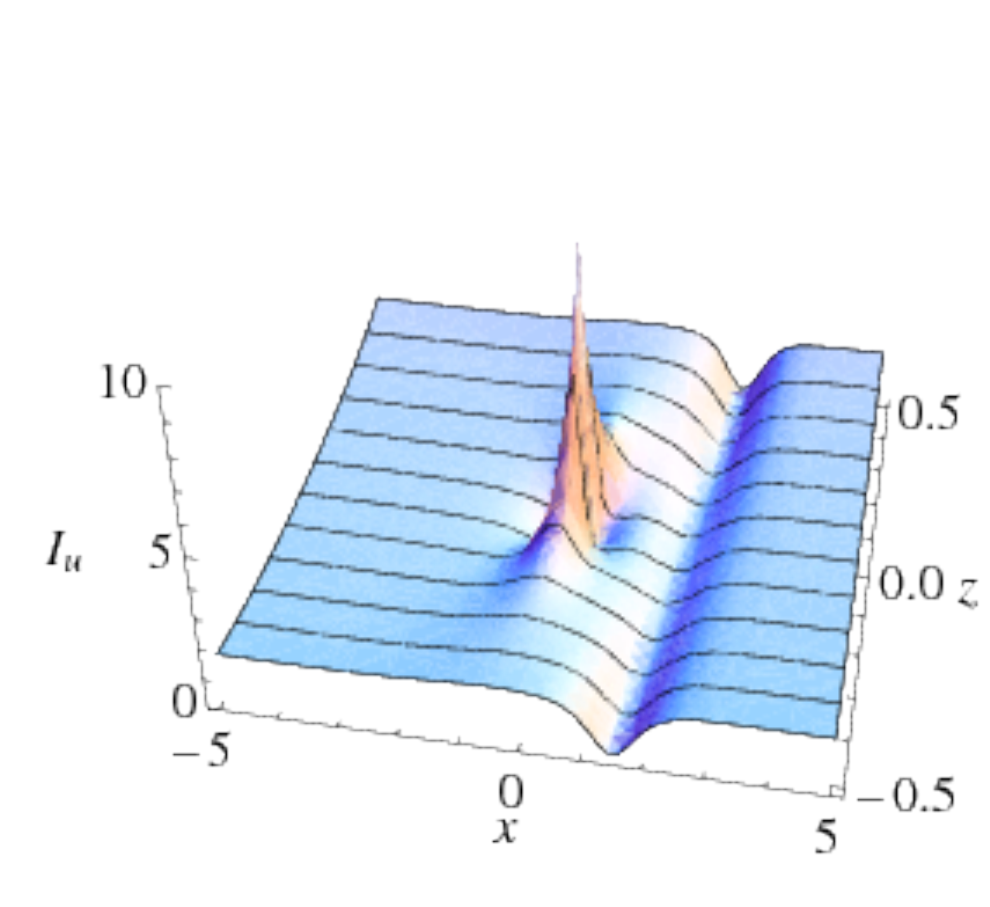}\includegraphics[width=0.5\textwidth,natwidth=310,natheight=242]{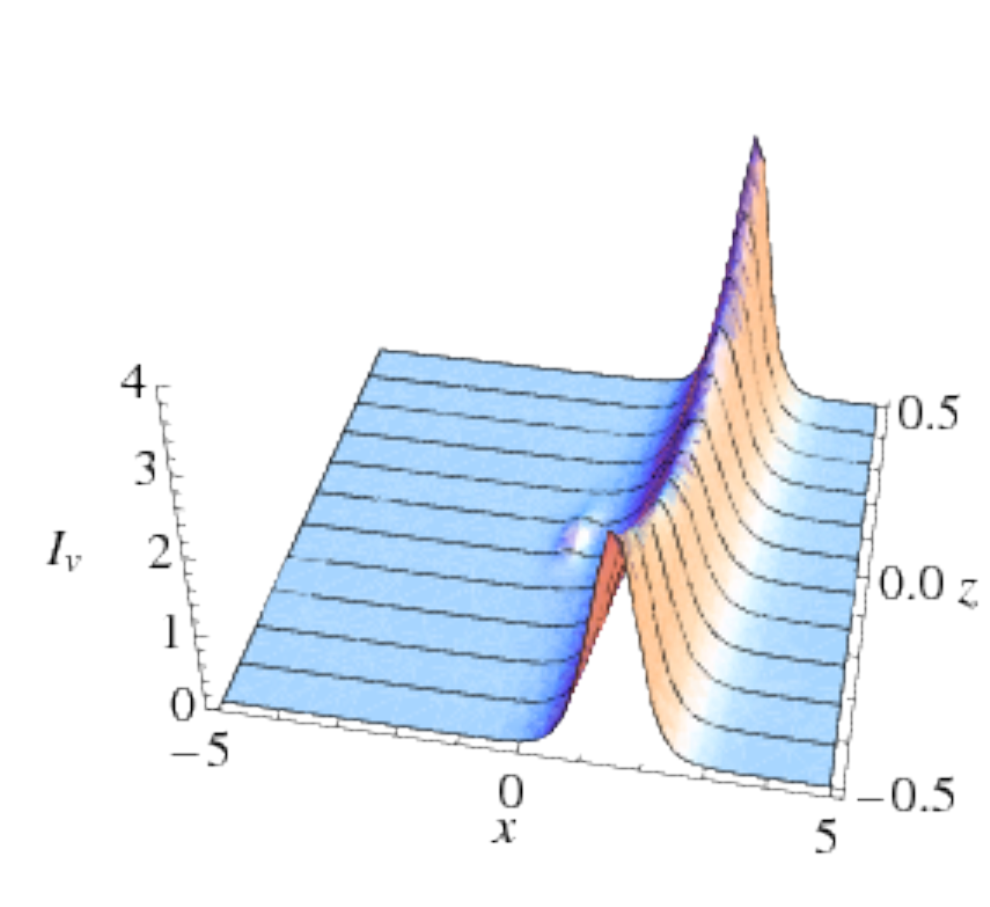}\\
\includegraphics[width=0.5\textwidth,natwidth=310,natheight=242]{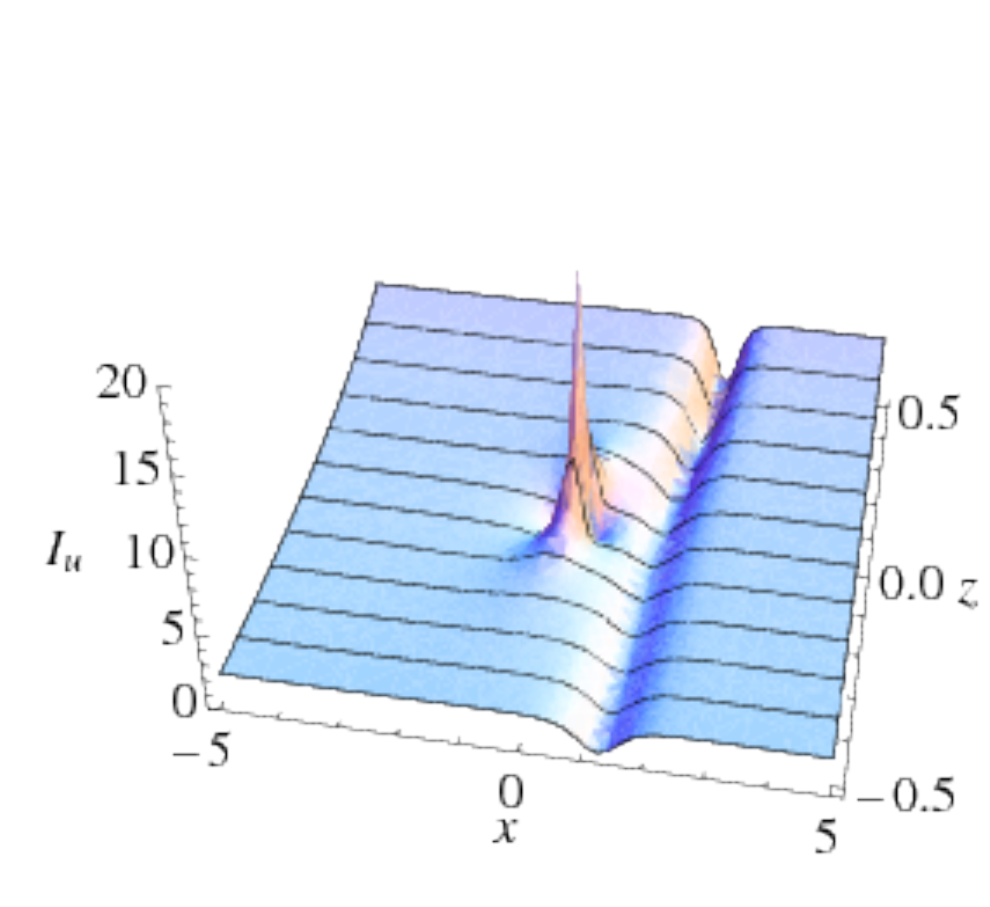}\includegraphics[width=0.5\textwidth,natwidth=310,natheight=242]{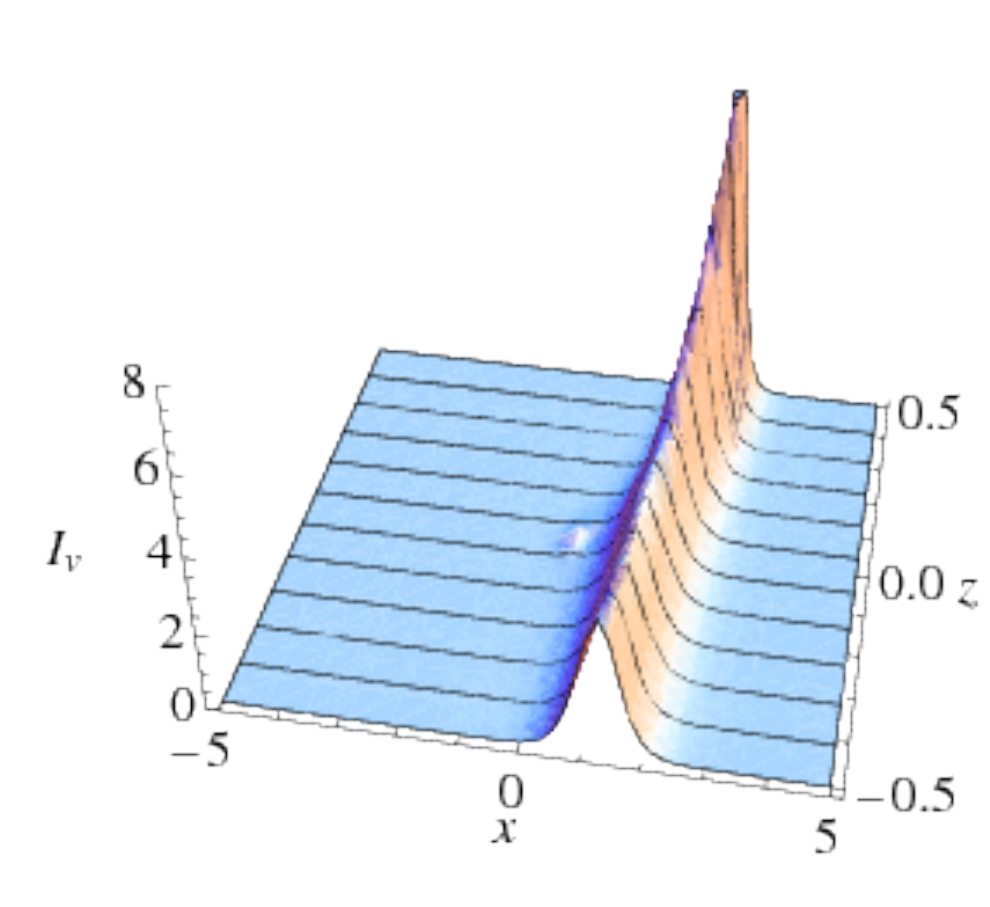}\\
\includegraphics[width=0.5\textwidth,natwidth=310,natheight=242]{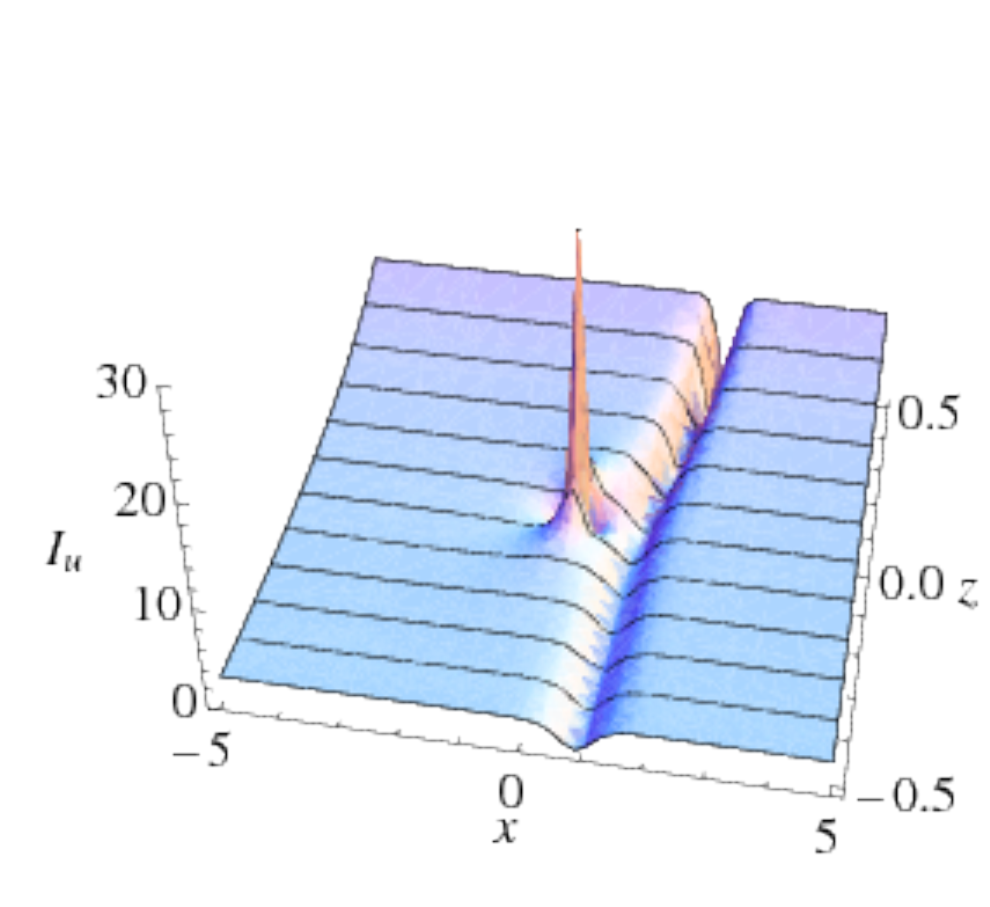}\includegraphics[width=0.5\textwidth,natwidth=310,natheight=242]{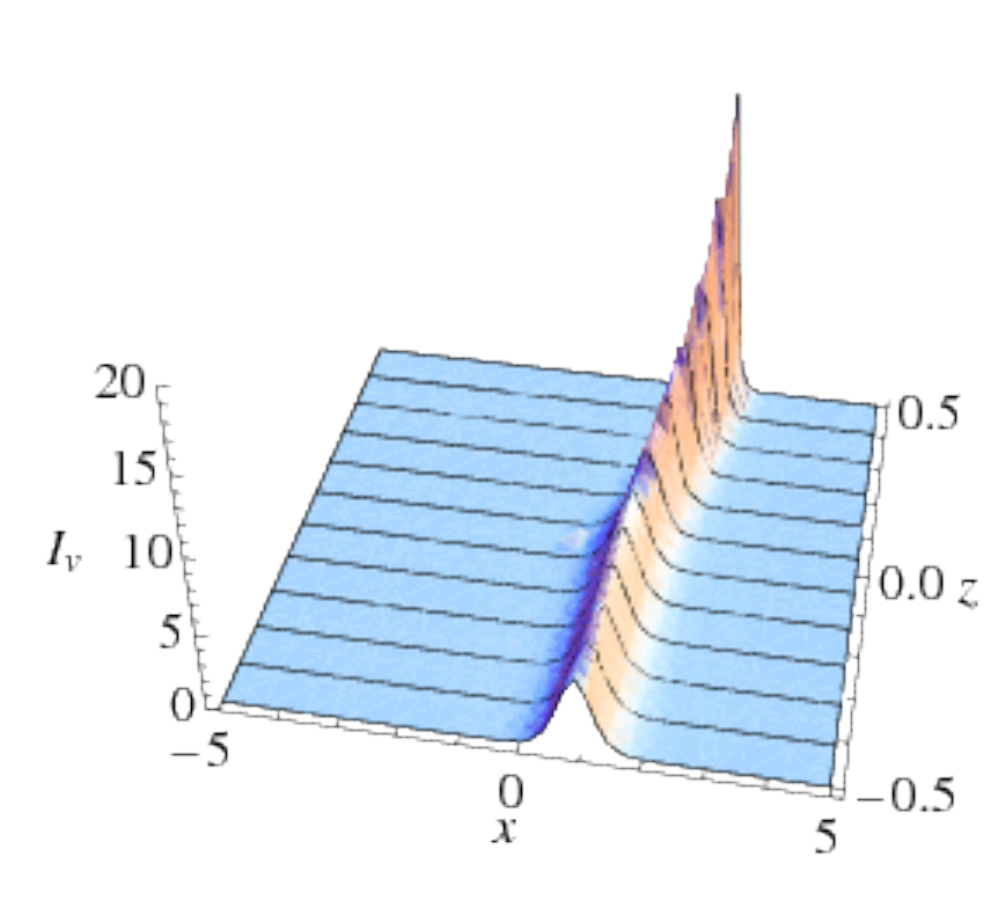}
\caption{\label{fig6}Freak wave intensity patterns for generalized
tapering. The wave parameters are $a_{1}=1, a_{2}=0, h=0.1$.
Riccati parameter $c=10, 1$, and $0.4$ for plots at the top,
middle and bottom. }
\end{center}
\end{figure}

We shall now discuss the effects of tapering on the intensity of
the freak waves represented by Equation (\ref{19}). Based on the
results of supersymmetric quantum mechanics, we generate a class
of tapering function and width as \cite{Kumar,Goyal} \be\label{20}
\widehat{f}(z)=f(z)-\frac{2}{\beta}\frac{d}{dz}\left(\frac{W^2(z)}{c+\int_{-\infty}^ZW^2(s)ds}\right),
\ee \be\label{21}
\widehat{W}(z)=\frac{\sqrt{c(c+1)}~W(z)}{c+\int_{-\infty}^ZW^2(s)ds},
\ee
  where $c$ is an integration constant, known as Riccati
parameter and $c \notin  [-2, 0]$. $\widehat{f}(z)$ is called
generalized tapering function and $\widehat{W}(z)$, the
generalized width. We shall consider $\sech^2$-type tapering
commonly used in fiber optics. As described in section II a
compatible pair of tapering function and width are \be\label{22}
f(z)= \frac{1}{\beta}[1-2~\mbox{sech}^2z]~ \mbox{and}~ W(z)=
\mbox{sech}~z.\ee It is now straight forward to obtain a class of
$\widehat{f}(z)$~ and~ $\widehat{W}(z)$ from Equation (\ref{20})
and (\ref{21}) as
\bea\label{23}\widehat{f}(z)=f(z)-\frac{2}{\beta}\frac{d}{dz}\left(\frac{\mbox{sech}^2z}{c
+ (1 + \tanh{z})}\right)\eea~and~\be\label{24}\widehat{W}(z)=
\frac{\sqrt{c(c+1)}~ \sech z}{c + (1 + \tanh{z})}.\ee Figure
\ref{fig6} depicts the intensity variation with generalized width.
We observe that, decrease in Riccati parameter value results in
increase of intensity. Another important observation is that, the
distance between the Peregrine soliton and dark/bright solitons
increases with the increasing value of Riccati parameter.
\section{Symbiotic similaritons} 
\begin{figure}[h*]
\includegraphics[scale=0.7]{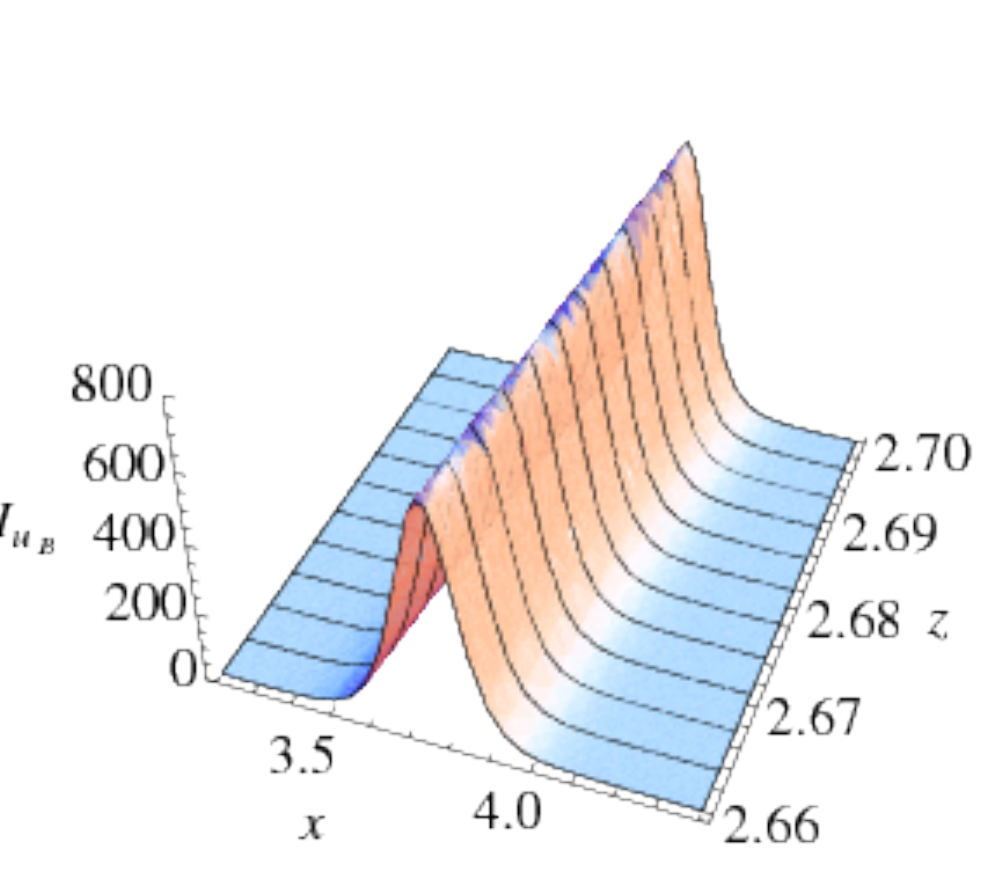}~~~~~\includegraphics[scale=0.7]{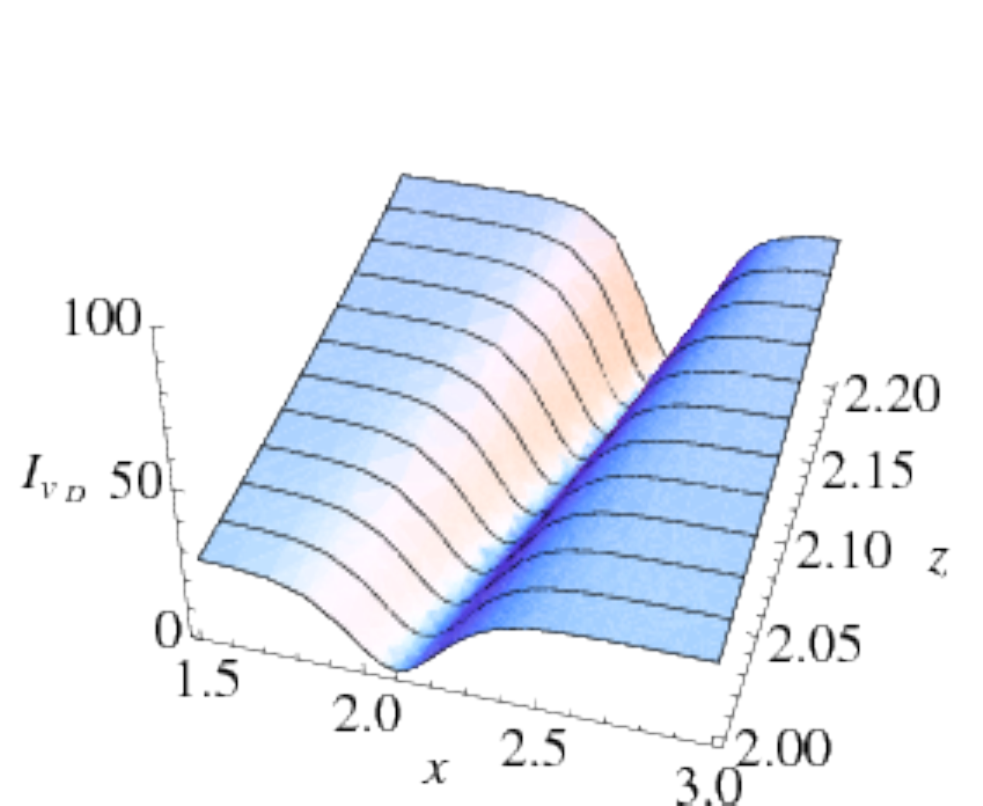}
\caption{\label{fig7}Intensity profiles of bright and dark
solitons with $r_{11}=5, r_{12}=6, r_{21}=6, r_{22}=8$ and $p=.5.$
}\end{figure}

Equation (\ref{14}) and (\ref{15})
may possess different soliton pairs as solutions. These soliton
pairs may appropriately be called as the symbiotic solitons;
because each wave in the pair depends crucially on the cross-phase
modulation from the complementary wave \cite{Lisak,Kivsher}. In
general, both solitons are interdependent on each other.\\
Assuming $\Psi_{1}= p~ e^{i\rho \zeta }~\mbox{sech}[\eta X]$ and
$\Psi_{2}= q~ e^{i\omega \zeta } \tanh[\eta X]$ as the general
form of the bright-dark soliton solutions of Equation (\ref{14})
and (\ref{15}); we have \cite{Ablowitz} \be\label{25}\Psi_{1}= p~
e^{i\frac{(r_{11}r_{22}-2r_{11}r_{12}+r_{12}r_{21})}{2(r_{22}-r_{12})}
p^{2}\zeta }~
\mbox{sech}\left[\sqrt{\frac{(r_{11}r_{22}-r_{12}r_{21})}{(r_{22}-r_{12})}}~
p X\right]\ee and \be\label{26}\Psi_{2}=
p~\sqrt{\frac{r_{21}-r_{11}}{r_{22}-r_{12}}}~
e^{i\frac{r_{22}(r_{21}-r_{11})}{(r_{22}-r_{12})} p^{2}\zeta }~
\tanh
\left[\sqrt{\frac{(r_{11}r_{22}-r_{12}r_{21})}{(r_{22}-r_{12})}}~
p X\right],\ee where $p$ is any arbitrary real constant. The
expressions for intensity of the bright-dark similariton solutions
can be written as
 \begin{align}\label{27}
I_{u_{B}}(x,y,z)&=
\frac{A_{0}^{2}}{W^2}p^{2}e^{2\Omega}~\mbox{sech}^2\left[\sqrt{\frac{(r_{11}r_{22}-r_{12}r_{21})}{(r_{22}-r_{12})}}
~p X\right],\\ \label{28} I_{v_{D}}(x,y,z)&=
\frac{A_{0}^{2}}{W^2}p^{2}e^{2\Omega}\frac{(r_{21}-r_{11})}{(r_{22}-r_{12})}~\tanh^{2}\left[\sqrt{\frac{(r_{11}r_{22}-r_{12}r_{21})}{(r_{22}-r_{12})}}
~p X\right] .\end{align}In Figure \ref{fig7} we have plotted
intensity of bright-dark soliton pair, corresponding to Equation
(\ref{27}) and (\ref{28}) for a set of values for coupling
coefficients and parameter $p$.
 Similarly, we have bright-bright soliton solutions
\be\label{29}\Psi_{1}= p~
e^{i\frac{(r_{11}r_{22}-r_{12}r_{21})}{2(r_{22}-r_{12})}
p^{2}\zeta }~
\mbox{sech}\left[\sqrt{\frac{(r_{11}r_{22}-r_{12}r_{21})}{(r_{22}-r_{12})}}~
p X\right],\ee \be\label{30}\Psi_{2}=
p~\sqrt{\frac{r_{11}-r_{21}}{r_{22}-r_{12}}}~
e^{i\frac{(r_{11}r_{22}-r_{12}r_{21})}{2(r_{22}-r_{12})}
p^{2}\zeta }~ \mbox{sech}
\left[\sqrt{\frac{(r_{11}r_{22}-r_{12}r_{21})}{(r_{22}-r_{12})}}~
p X\right]\ee and dark-dark soliton solutions
\be\label{31}\Psi_{1}= p~
e^{i\frac{(r_{11}r_{22}-r_{12}r_{21})}{(r_{22}-r_{12})} p^{2}\zeta
}~ \tanh
\left[\sqrt{\frac{(r_{12}r_{21}-r_{11}r_{22})}{(r_{22}-r_{12})}}~
p X\right],\ee \be\label{32}\Psi_{2}=
p~\sqrt{\frac{r_{11}-r_{21}}{r_{22}-r_{12}}}~
e^{i\frac{(r_{11}r_{22}-r_{12}r_{21})}{(r_{22}-r_{12})} p^{2}\zeta
}~ \tanh
\left[\sqrt{\frac{(r_{12}r_{21}-r_{11}r_{22})}{(r_{22}-r_{12})}}~
p X\right].\ee  Such types of solution pairs have been observed
experimentally in photo-refractive materials
\cite{Christodoulides} and also in the multi-component BEC in the
harmonic trap \cite{Yan}. Observing these solution pairs it is
clear that, the amplitude and phase of the propagating wave
components can be controlled by varying the strength of the SPM
and XPM coefficients. Also, due to different feasibility
conditions, these solution pairs cannot exist simultaneously.\\

Finally, we shall discuss about a special case where the two wave
components are directly proportional to each other.

\begin{figure}[h*]
\includegraphics[scale=0.7]{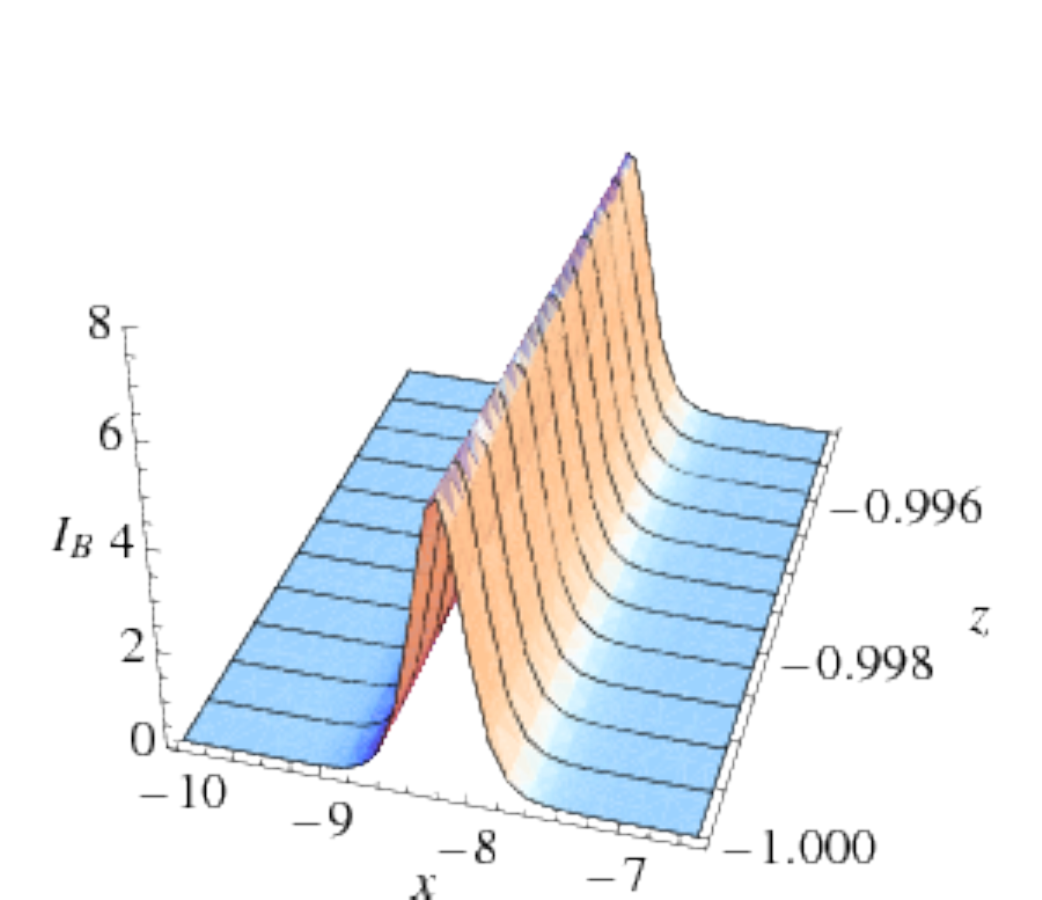}~~~~~\includegraphics[scale=0.7]{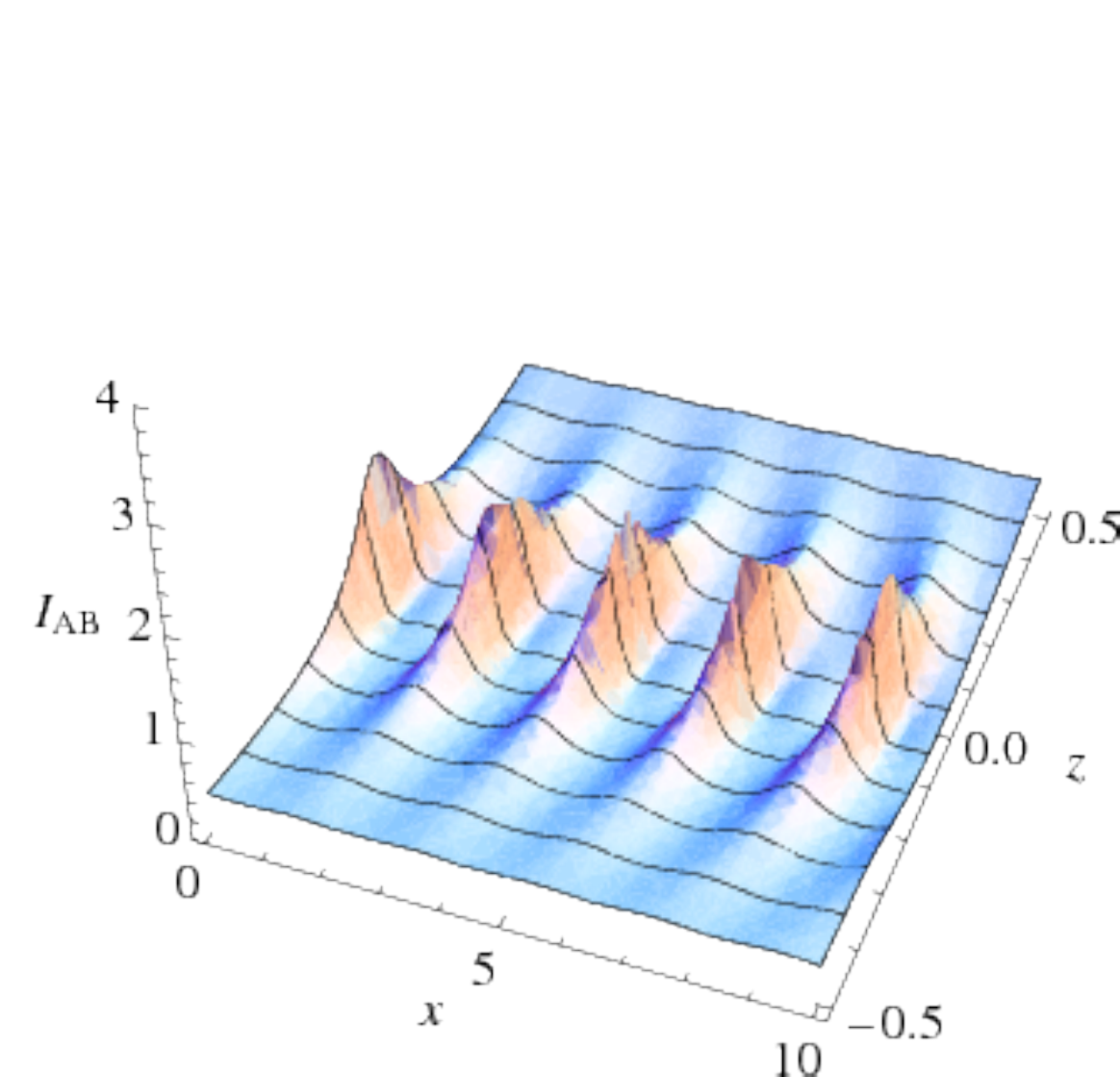}
\includegraphics[scale=0.7]{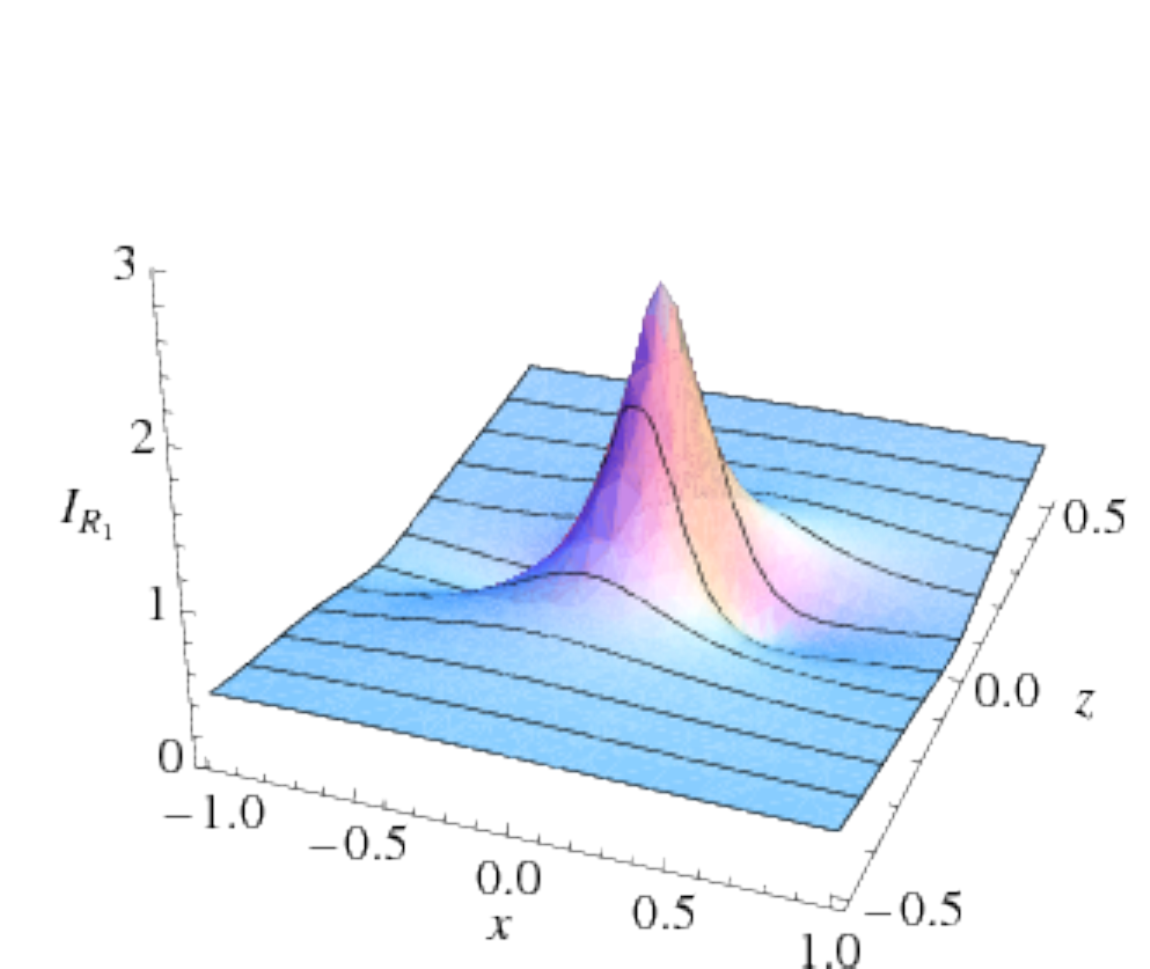}~~~~~\includegraphics[scale=0.7]{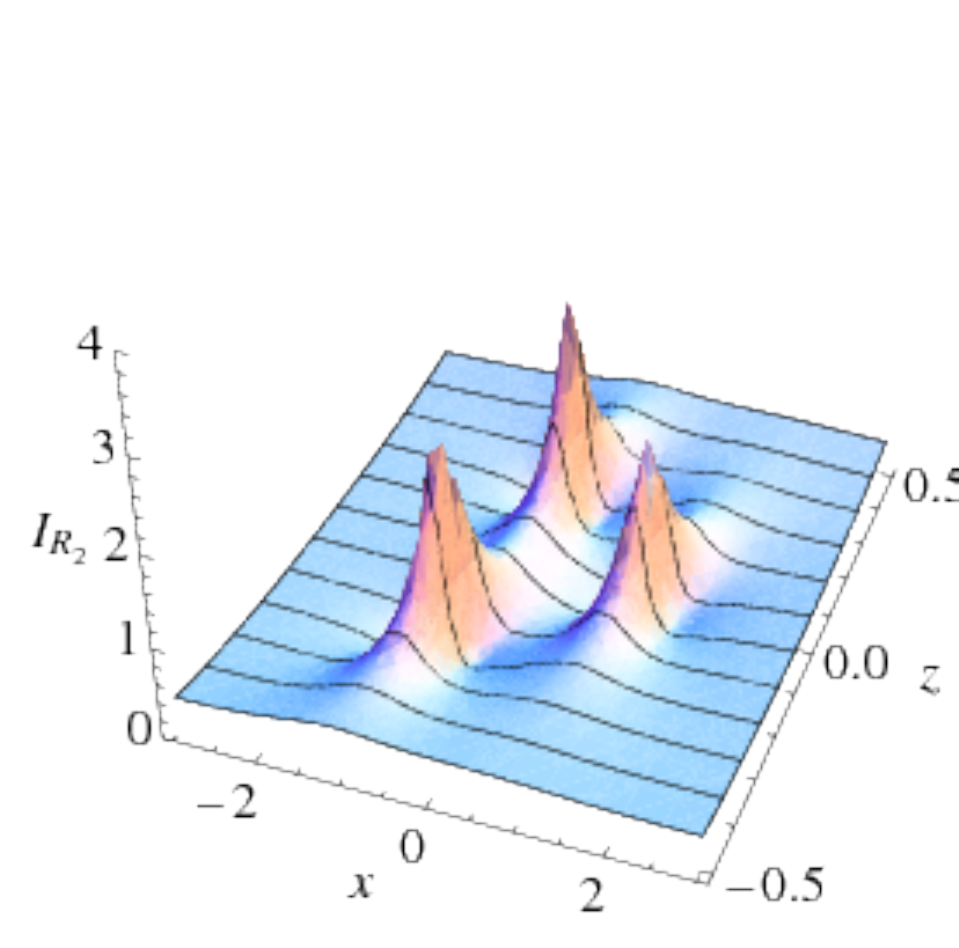}
\caption{\label{fig8}Intensity profiles of self-similar bright
soliton, Akhmediev breather, first-order rogue wave and
second-order rogue wave, respectively, with $r_{11}=5, r_{12}=6,
r_{21}=6, r_{22}=8$ and $v=2$.}
\end{figure}

\subsection{Proportional similaritons, breathers and rogons
} For $\Psi_{2} = \sqrt{\frac{r_{21}-r_{11}}{r_{12}-r_{22}}}~
\Psi_{1}$ and $\Psi_{1} \equiv  \Psi$ both Equation (\ref{14}) and
(\ref{15}) reduces to
\be\label{33}i\frac{\partial\Psi}{\partial\zeta}+\frac{1}{2}\frac{\partial^{2}\Psi}{\partial
X^{2}}+ \kappa |\Psi|^{2}\Psi=0\ee where $\kappa=\frac{r_{12}
r_{21} - r_{11} r_{22}}{r_{12}-r_{22}}$.
\\Now, assuming the general form of the bright soliton as $\mbox{sech} [p(X-v
\zeta)] \exp[i q (X-\omega \zeta)]$; we have
\be\label{34}\Psi_{B}= \mbox{sech} [\sqrt{\kappa}(X-v \zeta)]~
e^{i v (X-\frac{v^{2}-\kappa}{2v} \zeta)},\ee the bright soliton
solution of Equation (\ref{33}) where $v$ is the velocity of the
bright soliton.\\ Similarly, assuming the general form of the
fundamental dark soliton as $(q\tanh [q(X-v \zeta)] + i~p) \exp(i
\omega \zeta)$; we have for $\kappa=-1$ i.e. for
$r_{12}(r_{21}+1)=r_{22}(r_{11}+1)$, \be\label{35}\Psi_{D}= [q
\tanh [q(X-v \zeta)]+ i~v ]~ e^{ -i(v^{2}+q^{2})\zeta},\ee the
dark soliton solution of Equation (\ref{33}). Here $v$ is the
velocity of
dark soliton and $q$ is an arbitrary real constant.\\
For the focussing nonlinearity, Equation (\ref{33}) has another
localized solution called Akhmediev breathers given by
\be\label{36}\Psi_{AB}=
\frac{1}{\sqrt{\kappa}}\frac{\cos(\sqrt{2}X)+i \sqrt{2} \sinh
\zeta}{\cos(\sqrt{2}X)- \sqrt{2} \cosh \zeta}e^{i\zeta}\ee The
breathers are nonlinear waves which can carry energy in a
localized and oscillatory fashion. In contrast to solitons which
are localized in $X$, the breathers are localized in $\zeta$ and
oscillating in $X$. The breathers are not only a mathematical
concept but they have also been realized experimentally in
different systems such as in BEC \cite{Trombettoni}, dispersion
managed optical waveguides and fibers \cite{Kutz} and Josephson
arrays \cite{Trias}. Breather solutions have been obtained for
various nonlinear evolution equations like KdV \cite{Chow},
Gardener equation \cite{Slyunyaev}, modified-KdV \cite{Lamb} and
NLSE. In the context of NLSE, these spatially periodic solutions
are termed as Akhmediev breathers.\\ Equation (\ref{33}) has rogue
wave solution with the following basic structure:

 \be \label{37}\Psi_{R_{\nu}}= \frac{1}{\sqrt{\kappa}}\left[(-1)^{\nu}+\frac{H_{\nu}(X,\zeta)+ i K_{\nu}(X,\zeta))}{D_{\nu}(X,\zeta)}\right]e^{i\zeta},\ee
 where $\nu$ is the order of the solution; $H_{\nu},K_{\nu},D_{\nu}$ are polynomials of different order. For first order rogue wave solution
 \be\label{38} H_{1}=4,\ee \be\label{39} K_{1}=8\zeta,\ee \be\label{40} D_{1}=1+4\zeta^2+4 X^2 .\ee
 The $2$nd order rogue wave solution is the $\nu=2$ version of Equation (\ref{37}),
 with
 \be\label{41} H_{2}= 12[3-16X^4-24X^2(4\zeta^2+1)-4\alpha X-80\zeta^4-72\zeta^2+4\gamma \zeta],\ee
    \be\label{42}K_{2}= 24[\zeta(15-16X^4+24X^2-4\alpha X)-8(4X^2+1)\zeta^3-16\zeta^5+\gamma
  (2\zeta^2-2X^2-\frac{1}{2})],\ee and
  \bea D_{2}=64X^6+ 48 X^4(4\zeta^2+1)+12X^2(3-4\zeta^2)^2+64\zeta^6+432\zeta^4+396\zeta^2+9 \no \\+\alpha [\alpha +4X(12\zeta^2-4X^2+3)]+\gamma [\gamma +4\zeta(12
  X^2-4\zeta^2-9)]\label{43},\eea
  where $\alpha$ and $\gamma$ are arbitrary real constants. Rogue wave, sometimes known as freak wave or monster wave is briefly formed, single, exceptionally large
amplitude wave. Rogue waves are localized both in $X$ and $\zeta$.
These waves are of great interest in a variety of complex systems,
from optics and fluid dynamics to Bose-Einstein condensates and
finance. We should mention here that, just as solitary waves are
known as solitons, the term rogons is coined for rogue waves. The
corresponding terminologies `oceanic rogons', `optical rogons' and
`matter rogons' are used in the field of hydrodynamics, nonlinear
optics and BECs, respectively.\\ Now, $I_{v}=a I_{u}$ where
$a=\frac{r_{21}-r_{11}}{r_{12}-r_{22}}$. Because of the constant
scaling factor the intensity $I_{v}$ will have same profile as
$I_{u}$ for $a>0$. Therefore, in this section we shall draw
intensity diagrams associated to $I_{u}$ only. The general
expression
  of the intensity for bright similaritons $I_B$, self-similar Akhmediev breathers $I_{AB}$, self-similar first-order rogue waves $I_{R_1}$ and
   self-similar second-order rogue waves $I_{R_2}$ are given
  as
 \begin{align}\label{44}
I_B(x,y,z)&=\frac{A_{0}^{2} e^{2\Omega}}{W^2}\mbox{sech}^2[\sqrt{\kappa}(X-v\zeta)],\\
\label{45}I_{AB}(x,y,z)&=\frac{A_{0}^{2} e^{2\Omega}}{\kappa
W^2}\left[\frac{\cos^2(\sqrt{2}X)+2~\sinh^2\zeta}{\left(\cos(\sqrt{2}X)-\sqrt{2}\cosh\zeta\right)^2}
\right],\\
\label{46}I_{R_1}(x,y,z)&=\frac{A_{0}^{2} e^{2\Omega}}{\kappa
W^2}\left[ 1 + 8 \frac{1 + 4
\zeta^2 - 4 X^2}{(1 + 4 \zeta^2 + 4 X^2)^2}\right],\\
\label{47}I_{R_2}(x,y,z)&=\frac{A_{0}^{2} e^{2\Omega}}{\kappa W^2}
\left[\frac{K_{2}^{2}+(H_{2}+
D_{2})^{2}}{D_{2}^2}\right].\end{align} In Figure \ref{fig8} we
have shown intensity profiles of self-similar bright soliton,
Akhmediev breather, first-order rogue wave and second-order rogue
wave corresponding to Equation (\ref{44})-(\ref{47}),
respectively, for a fixed set values for coupling coefficients and
velocity.

\begin{figure}[h*]
\includegraphics[scale=0.7]{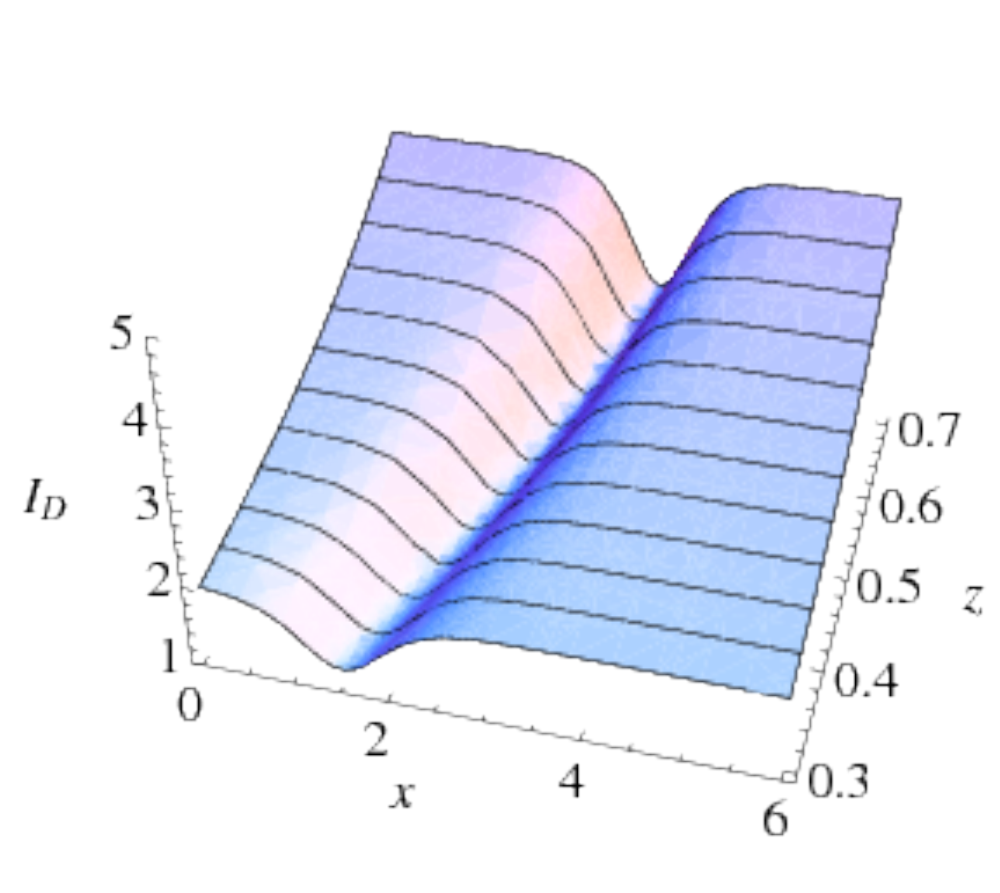}
\caption{\label{fig9}Intensity profile of self-similar dark
soliton, with $v=1$ and $q=0.8$.}
\end{figure}

 For $\kappa=-1$, we have dark soliton
solution represented by Equation (\ref{35}). Hence, the general
expression of the intensity for dark similaritons $I_D$ is \bea
I_D(x,y,z)=\frac{A_{0}^{2} e^{2\Omega}}{W^2}(v^{2}+
q^{2}\tanh^{2}[q(X-v \zeta)])\label{48}.\eea In Figure \ref{fig9}
we have plotted intensity of self-similar dark soliton,
corresponding to Equation (\ref{48}) for particular values of
parameters $v$ and $q$.

\section{Conclusion}We have used similarity transformations to
convert vc-CNLSE into constant co-efficient CNLSE. For the
associated Manakov system, we could extract a novel
multi-parametric solution having both exponential and rational
dependence on coordinates, under certain functional constraints.
This novel solution generate a family of solutions for different
parameter values; e.g. Peregrine soliton (rational), combination
of either bright soliton and rogue wave or dark soliton and rogue
wave or breather soliton and rogue wave. We have also shown the
effect of generalized tappering on the intensity of these
semirational waves by changing Riccati parameter. Further, we
could establish symbiotic existence of different soliton pairs as
solutions for a set of self-phase modulation (SPM) and cross-phase
modulation (XPM) coefficients. These solution pairs may consists
of one bright and one dark soliton, two bright solitons or two
dark solitons. Finally, assuming direct proportionality between
wave components, we could extract bright and dark similaritons,
self-similar breathers and rogue waves as solutions. Because of
the universality of the vc-CNLSE, these solutions may contribute
to better control and understanding of various solitary wave
phenomena in a variety of complex dynamics, ranging from fluid
dynamics, to optical communications, Bose-Einstein condensates,
and financial systems.\\For future studies, one can investigate
the effect of gain on nonlinearity and the wave intensity. Studies
regarding stability of the different solution is also an open
problem.
\section*{Acknowledgement} T.S.R. acknowledges support from the
DST, Government of India, through Fast Track Project Ref.
No.--SR/FTP/PS-132/2012, during the course of this work.

\end{document}